\documentclass[prl,showpacs,superscriptaddress,preprintnumbers,amssymb,twocolumn]{revtex4}
\usepackage{graphicx}
\usepackage{amssymb}
\begin{document}
\input epsf
\def\beq{\begin{equation}}
\def\eeq{\end{equation}}
\def\beqn{\begin{eqnarray}}
\def\eeqn{\end{eqnarray}}

\title{Topological invariants of time-reversal-invariant band structures}

\author{J.~E.~Moore}
\affiliation{Department of Physics, University of California,
Berkeley, CA 94720} \affiliation{Materials Sciences Division,
Lawrence Berkeley National Laboratory, Berkeley, CA 94720}
\author{L.~Balents}
\affiliation{Department of Physics, University of California,
Santa Barbara, CA 93106}
\date{\today}

\begin{abstract}
  The topological invariants of a time-reversal-invariant band
  structure in two dimensions are multiple copies of the
  $\mathbb{Z}_2$ invariant found by Kane and Mele.  Such invariants
  protect the topological insulator and give rise to a spin Hall
  effect carried by edge states.  Each pair of bands related by time
  reversal is described by a single $\mathbb{Z}_2$ invariant, up to
  one less than half the dimension of the Bloch Hamiltonians.  In
  three dimensions, there are four such invariants per band.  The
  $\mathbb{Z}_2$ invariants of a crystal determine the transitions
  between ordinary and topological insulators as its bands are
  occupied by electrons.  We derive these invariants using maps from
  the Brillouin zone to the space of Bloch Hamiltonians and clarify
  the connections between $\mathbb{Z}_2$ invariants, the integer
  invariants that underlie the integer quantum Hall effect, and
  previous invariants of ${\cal T}$-invariant Fermi
  systems.%The $\mathbb{Z}_2$ topological invariant of
          %time-reversal-invariant (${\cal T}$-invariant) fermion band
          %structures found by Kane and Mele underlies the topological
          %insulator, a new phase of matter.  This invariant is first
          %derived using well-known results from homotopy theory,
          %which clarifies the relationship between this invariant and
          %previous work on integer-valued topological invariants in
          %${\cal T}$-invariant Fermi systems and the integer quantum Hall
          %effect.  It is then generalized to multiple bands: there is
          %generically a family of independent $\mathbb{Z}_2$
          %invariants, with one independent invariant per pair of
          %${\cal T}$-conjugate bands when the Bloch Hamiltonian Hilbert
          %space is infinite-dimensional and there are no additional
          %symmetries.  The maximal number of invariants is one less
          %than half the dimension of this Hilbert space.  These bulk
          %invariants combine into a single $\mathbb{Z}_2$ invariant
          %in the presence of an edge, because of degeneracies.
\end{abstract}
\pacs{73.43.-f, 85.75.-d, 73.20.At}
\maketitle

In a remarkable pair of papers, Kane and Mele~\cite{km1,km2} proposed
a $\mathbb{Z}_2$ topological invariant of time-reversal-invariant
insulators in two dimensions, showed that the nontrivial ``topological insulator''
phase created by spin-orbit coupling has an intrinsic spin Hall effect distinct from earlier
proposals~\cite{murakami,sinova}, and argued that graphene is a viable
system in which to observe this effect.  (Here $\mathbb{Z}_2 \equiv
\mathbb{Z} / 2 \mathbb{Z}$ is the cyclic group of two elements.)  A
direct and experimentally relevant characterization of the invariant
was given in terms of edge states at the boundary of a 2D insulator:
the topological insulator has an odd number of Kramers pairs of edge
modes, while the ordinary insulator has an even number.  Two explanations for the invariant as a property of the bulk band structure with spin-orbit coupling were also given.

We first show that time-reversal-invariant (${\cal T}$-invariant) 2D insulators
have multiple
$\mathbb{Z}_2$ invariants that are directly analogous to the band TKNN
integers or Chern numbers in the integer quantum Hall effect (IQHE)
and can be understood using similar methods.  This clarifies and
expands upon the two original descriptions of a single bulk
$\mathbb{Z}_2$ invariant, which do not follow the standard
homotopy paradigm of most topological invariants in
condensed matter physics, and whose connection to the IQHE is opaque:
for the case of an occupied pair of bands relevant to graphene, the
$\mathbb{Z}_2$ invariant was explained either by invoking K-theory
(recently used to classify Fermi surfaces~\cite{horava}) or by
counting zeroes or zero regions of a certain matrix Pfaffian.
%We find that there are multiple $\mathbb{Z}_2$ invariants in 2D and discuss  relationship to integer quantum Hall physics, and their generalization to higher dimensions and interacting systems are the main results of this paper.

The $\mathbb{Z}_2$ invariant is first rederived for the simplest case
(two occupied bands related by time-reversal symmetry) using basic
homotopy theory to establish its connection to the usual Chern number or
TKNN integer~\cite{tknn} of ${\cal T}$-invariant systems.  There is a
connection between its absence with inversion symmetry and previous work
on topological invariants of ${\cal T}$-invariant Fermi systems~\cite{asss}.
We then obtain results for multiple occupied bands: there are multiple
$\mathbb{Z}_2$ invariants in such systems, and perhaps most
interestingly, four invariants per band in 3D.  In 2D, the
significance of these invariants in physical systems is straightforward:
just as two IQHE states with the same {\sl sum} of Chern numbers for
occupied bands are adiabatically connected, two ${\cal T}$-invariant band
insulators are adiabatically connected if and only if they have the same
$\mathbb{Z}_2$ sum of individual $\mathbb{Z}_2$ invariants.

%Just as a phase in the integer quantum Hall effect is determined by the sum of band Chern numbers for occupied bands, the phase of a ${\cal T}$-invariant insulator is fixed by the $\mathbb{Z}_2$ sum of the $\mathbb{Z}_2$ invariants for occupied bands.
%with no accidental degeneracies, i.e., the single-$\mathbb{Z}_2$ description is valid but incomplete.

The approach in this paper can be outlined as follows.  The basic objects of homotopy theory are the homotopy groups $\pi_n(M)$ that describe equivalence classes under smooth deformations of mappings from the sphere $S^n$ to a manifold $M$.  A band structure can be thought of as a map from the Brillouin zone (a torus rather than a sphere) to the space of Bloch Hamiltonians, assuming that the Hilbert space is the same for all points in the Brillouin zone.  The ${\cal T}$ symmetry means that the effective Brillouin zone (EBZ), a set of points for which the Bloch Hamiltonians can be specified independently, is nearly a sphere: we consider ``contractions'' that extend a mapping from the EBZ to one from the sphere.

A given mapping from the EBZ has an infinite set of topologically inequivalent contractions in this sense.  We consider a mapping from the EBZ together with the set of all of its contractions, which is invariant under smooth deformations of the original mapping.  There are then exactly two equivalence classes, for the case of two occupied bands connected by ${\cal T}$, which correspond to the ordinary insulator and the topological insulator.  In this simplest case, an ordinary (topological) insulator becomes the equivalence class of all mappings from the sphere with even (odd) Chern numbers.  This construction requires only standard homotopy results, makes no assumptions about the details of the band structure or the existence of additional commuting operators such as spin, and generalizes directly to multiple bands and higher dimensions.

For 2D ${\cal T}$-breaking systems, in a nondegenerate band structure each band is associated with a TKNN integer that is invariant under smooth perturbations of the Bloch Hamiltonians.  More precisely, if the Hilbert space of the Bloch Hamiltonians is finite and of dimension $n$, there are $n-1$ independent integer-valued invariants~\cite{ass} because the invariants sum to zero.   There are a number of ways to obtain the TKNN integers, e.g., as an integral over the magnetic Brillouin zone (a torus) of the Berry flux, using the explicit wavefunction $\psi_i$ for band $i$, or as an integral using the projection operator~\cite{ass}
\beq
n_i = {i \over 2 \pi} \int_{BZ}\,{\rm Tr}\,(dP_i\,P_i\,dP_i ), \quad P_i = |\psi_i \rangle \langle \psi_i |.
\label{tknn}
\eeq
Here $dP_i = dx \partial_x P_i + dy \partial_y P_i$ and $dx dy = - dy dx$.  A powerful way to understand these integer invariants~\cite{ass} is by considering mappings of the torus to Bloch Hamiltonians, which are assumed to be nondegenerate and act upon the same Hilbert space.  Denoting the space of such Hamiltonians as $\cal M$, the existence of the TKNN integers follows from the first two homotopy groups
\beq
\pi_1(M) = 0, \pi_2(M) = \mathbb{Z}^{n-1},
\eeq
where the second formula indicates $n-1$ copies of the infinite cyclic group $\mathbb{Z}$.  The second result follows~\cite{ass} from the same exact sequence as used in the theory of topological defects~\cite{merminrev}.
For a pair of bands $i, j$ that are possibly degenerate with each other but with no other bands, there is a single integer-valued invariant~\cite{ass} obtained by replacing $P_i$ in (\ref{tknn}) with $P_{ij} = P_i + P_j$.

%~\cite{footnote-homotopy}

\begin{figure}[h!]
\epsfxsize = 3.25in
\epsffile{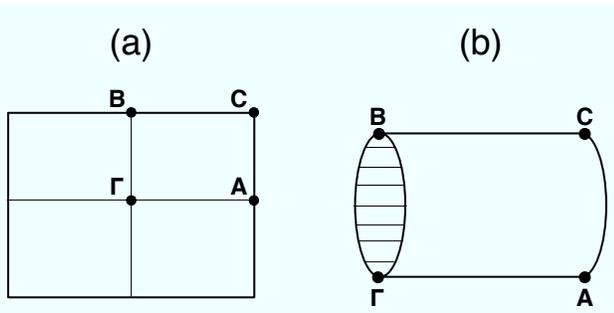} \caption{The topology of the effective Brillouin zone (EBZ): if the original Brillouin zone is the torus in (a), then ${\cal T}$-invariance reduces the independent degrees of freedom to live on the manifold in (b).  Points on the boundary circles that are connected by horizontal lines are conjugate under ${\cal T}$; the points $\Gamma, A, B, C$ are self-conjugate, and their Bloch Hamiltonians are therefore in the even subspace ${\cal Q}$. }
\end{figure}

Now consider consequences of invariance under the time-reversal operator ${\cal T}$.  For fermions, $T^2 = -1$ and ${\cal T}$ is represented by an antiunitary operator $\Theta$ in the Hilbert space of Bloch Hamiltonians.  Time-reversal connects both pairs of points in the Brillouin zone $(k,-k)$ and the associated Bloch Hamiltonians:
\beq
H(-k) = \Theta H(k) \Theta^{-1}.
\eeq
Fig.~1 shows the original toroidal Brillouin zone and the EBZ: specifying the Hamiltonians for all points on the EBZ determines them everywhere.  The Bloch Hamiltonians can be specified independently except at the boundaries, where points are related by ${\cal T}$.  Clearly points at which $k=-k$, such as $\Gamma, {\rm \bf A, B, C}$ in Fig.~1, are special: at these points the Bloch Hamiltonian commutes with $\Theta$.  This ``even'' subspace in the language of Ref.~\onlinecite{km1} has a natural interpretation as the set of symplectic or quaternionic Hamiltonians~\cite{asss}: we denote this subspace, with the additional assumption of no degeneracies other the two-fold degeneracies required by ${\cal T}$,  as ${\cal Q}$.  ${\cal T}$-invariance requires an even number of bands $2n$.

In general, a ${\cal T}$-invariant system need not have Bloch Hamiltonians in ${\cal Q}$ except at these special points as long as inversion symmetry is broken, so that $H(k) \not= H(-k)$.  At first glance, there is no obvious topological invariant for a degenerate band with time-reversal symmetry, because the Chern number for the whole Brillouin zone vanishes.  It is simplest to see this for the projection operator $P_i$ corresponding to a single nondegenerate band: using an explicit representation $P_i = |\psi_i\rangle \langle \psi_i |$,
\beqn
n_i &=& {i \over 2 \pi} \int_{BZ}\,{\rm Tr} (dP_{i}\,P_{i}\,dP_{i}) \cr
&=&
{i\over 2 \pi} {\rm Im} \int\,\Big[\langle \partial_x \psi_i | \partial_y \psi_i \rangle
%\cr &&
+ \langle \partial_x \Theta \psi_i | \partial_y \Theta \psi_i \rangle \Big] \cr
&=&
{i\over 2 \pi} {\rm Im} \int\,\Big[ \langle \partial_x \psi_i| \partial_y \psi_i \rangle
%\cr &&
+ \langle \partial_y  \psi_i | \partial_x \psi_i \rangle \Big] = 0.
\label{chernvanish}
\eeqn
A similar argument applies for two possibly degenerate bands, but there is nevertheless a topological invariant~\cite{km1}, and in general multiple topological invariants.

As a quick example of homotopy arguments, suppose that inversion symmetry is unbroken, which implies that the Bloch Hamiltonians are everywhere in ${\cal Q}$.  Any mapping of the torus $T^2$ to ${\cal Q}$ with the condition that points $k$ and $-k$ go to the same point is determined by its behavior on one point from each $(k,-k)$ pair, and the EBZ under this condition is topologically identical to a sphere (stitching ${\cal T}$-conjugate points together in Fig. 1b): the classes of such mappings are given by~\cite{asss} $\pi_2({\cal Q}) = 0$.  Hence there is no homotopy invariant for 2D band structures with both ${\cal T}$ and inversion symmetry, although $\pi_4({\cal Q}) \not = 0$ and higher-dimensional invariants can exist.

Now consider possible invariants without inversion symmetry.  We seek to classify mappings from $T^2$ to the space of Hamiltonians ${\cal C}$ that have at most two-fold degeneracies within a pair of ${\cal T}$-related bands and behave under $\Theta$ as specified above.
%More precisely, it is assumed that eigenvalues $1$ and $2$ are possibly degenerate with each other and always less than eigenvalues $3$ and $4$, which are again possibly degenerate and always less than eigenvalues $5$ and $6$, and so forth.
Such a mapping is determined by a mapping from the EBZ, i.e., a mapping from the cylinder $C$ to ${\cal C}$ with certain conditions on the two circular boundaries reflecting time-reversal.  The image of a boundary point must be the $\Theta$ conjugate of the image of the point on the same boundary related by $k \leftrightarrow -k$.

If the same Hamiltonian occurred at all points on the boundary, then the
topology of the cylinder becomes that of the sphere, and the degenerate
Chern numbers~\cite{ass} are integer-valued invariants for mappings from
the sphere to ${\cal C}$ that give one integer for each possibly
degenerate pair since $\pi_2({\cal C}) = \mathbb{Z}^{n-1}$.  For the
simplest case of two occupied possibly degenerate bands, the invariant
combination reduces to $n_1 + n_2$ if the bands are nondegenerate and
hence have separate Chern numbers $n_1$ and $n_2$.

We show first for the simplest case of one pair of bands that any
mapping from $C$ to ${\cal C}$, even if the elements at a boundary are
not all the same, can be smoothly deformed (``contracted'') to one in
which the boundary elements are identical to an arbitrary reference
element $Q_0 \in {\cal Q}$ (Fig.~2a); the resulting map from the {\it sphere}
has a well-defined Chern number.  It is required that at each stage of
the contraction, the boundary has the same conjugacy of points under
${\cal T}$ as in the original boundary.  This guarantees that two maps from
the EBZ that can be contracted to maps from the sphere with the same
Chern number are homotopic (deformable to each other).

%This requirement guarantees that given two maps from the EBZ that are deformable into each other, there are contractions of the two maps that give topologically equivalent mappings from the sphere.  In other words, the set of all mappings from the sphere generated by a mapping from the EBZ plus the set of all its contractions is a topological equivalence class, i.e., invariant under smooth deformations of the original map from the EBZ.

Then it is shown that different contractions differ by an arbitrary even
Chern number, so that there are only two invariant classes according to
whether the Chern number with any contraction is an odd or even integer.
The existence of at least one contraction follows immediately from
$\pi_1({\cal C}) = 0$: contract one side of the boundary circle on the
point $Q_0$, keeping the top and bottom points in ${\cal Q}$, and then
determine the other side's behavior by ${\cal T}$ conjugacy.  The fact that
different contractions differ by an even Chern number is less trivial
and gives an unexpected picture of how the $\mathbb{Z}_2$ invariant
arises: as an embedding of one $\mathbb{Z}$
homotopy group in another.  Whether a band is ``odd'' or ``even'' is
directly computable by using any contraction to define the Chern number
integral.

A contraction in the sense above gives a smooth mapping from a cylinder
to ${\cal C}$, $f(\theta,\lambda)$, $0 \leq \lambda \leq 1$ such
that for each $\lambda$, $f(\theta,\lambda)$ and $f(2 \pi -
\theta,\lambda)$ are ${\cal T}$ conjugates (which implies that $f(0,\lambda)$
and $f(\pi, \lambda)$ are in ${\cal Q}$).  Now $f(\theta,0)$ should
agree with the initial specification of Bloch Hamiltonians, while
$f(\theta,1)=Q_0$ is constant.  If both boundaries are contracted, the
resulting sphere has a well-defined Chern number for each pair of bands.

In fact, many topologically inequivalent contractions exist, and these
different contractions are responsible for reducing the integer-valued
invariants on the sphere to $\mathbb{Z}_2$ invariants on the EBZ.  Let
$f_1$ and $f_2$ be two different contractions.  Then define a
mapping $g(\theta,\lambda)$, which by composition changes from
contraction $f_1$ to $f_2$ (Fig.~2b): 
\beq
g(\theta,\lambda) = \cases{f_1(\theta,1-2 \lambda) &if $0 \leq \lambda < 1/2$ \cr
f_2(\theta,2 \lambda - 1) & if $1/2 \leq \lambda \leq 1$}.
\label{contcombine}
\eeq Although the domain of the mapping $g$ is topologically a sphere
because the circles at both ends go to the same point, $g$
differs in its ${\cal T}$ symmetry from the contracted half of the Brillouin 
zone, which in its interior has no ${\cal T}$ symmetry relating different values
of $\theta$.  Let the cylinder in the definition of $g$ have coordinates $\theta \in
[0,2\pi)$, $\lambda \in [0,1]$; then points $(\theta, \lambda)$ and $(2 \pi -
\theta,\lambda)$ are ${\cal T}$-conjugate.

\begin{figure}[h!]
\epsfxsize = 3.25in
\epsffile{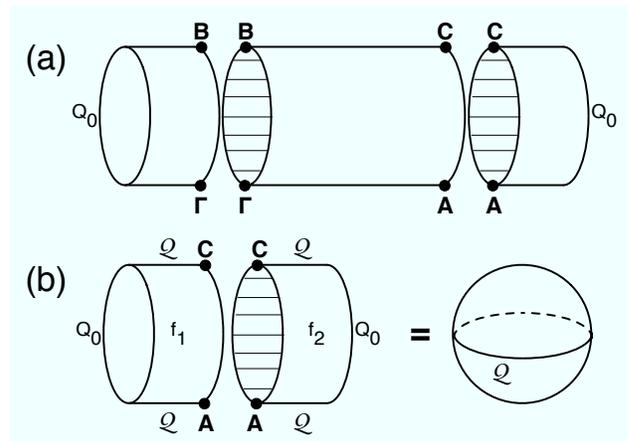} \caption{(a) Contracting the extended Brillouin
  zone to a sphere.  (b) Two contractions can be combined according to
  (\ref{contcombine}) to give a mapping from the sphere, but this
  sphere has a special property: points in the northern hemisphere are
  conjugate under ${\cal T}$ to those in the south, in such a way that
  overall the Chern number must be even.} 
\end{figure}

It follows that $g$ has arbitrary {\it even} Chern numbers, and the change in the Chern numbers of $n_B$ induced by changing from contraction $f_1$ to contraction $f_2$ is
\beq
\Delta n^i_B = 2 n^i_S,
\eeq
where $i$ indexes band pairs.  This can be verified in two steps: map the equator of the sphere $S$ to the constant element $E_0$, which is possible since $\pi_1({\cal Q}) = 0$ and topologically unique since $\pi_2({\cal Q}) = 0$, then note that each hemisphere has a well-defined Chern number and that the Chern numbers of the two hemispheres are {\it equal}, rather than {\it opposite} as in the case of the original Brillouin zone.  The reason for this equality can be understood easily in the cylindrical coordinates above, where the equator is at $\theta=\pi$ and $\theta=0$.  The identification under ${\cal T}$ of $\theta$ and $2 \pi - \theta$ means that $d\theta$ changes sign between a point and its time-reversal conjugate, but $d\lambda$ does not, giving an additional change of sign in equation (\ref{chernvanish}).

%Big question: how many invariants are there for n-1 or n/2 copies of Z2?  Reduction to simpler cases suggests n-1.

The same topological argument applies to $n$ pairs of bands.  Since
there is one integer invariant for each such pair with a zero sum rule, there is one
$\mathbb{Z}_2$ invariant for each pair, with an even number of ``odd'' band pairs.

We now give the generalization to three-dimensional Brillouin zones, where there are significant differences between $\mathbb{Z}_2$ invariants and the 3D integer-valued TKNN invariants~\cite{tknn,ass}: there are four independent $\mathbb{Z}_2$ invariants per pair of bands, even though there are only 3 Chern numbers for a pair of degenerate bands.  An alternate way to obtain the 2D result, which is more useful for 3D, is by considering contractions to a torus, rather than a sphere.
% since $\pi_1(C) = 0$, the classes of mappings from the torus to ${\cal C}$ are the same as those from the sphere, and again different contractions differ by an even Chern number.
The set of mappings from the Brillouin zone $T^3$ to ${\cal C}$ is determined by three ordinary Chern numbers since $\pi_3({\cal C}) = 0$: the three integers correspond to the $xy$, $yz$, and $xz$ planes~\cite{ass}.  The mappings from any two parallel planes in $T^3$ to ${\cal C}$ have the same Chern number.

Now consider possible $\mathbb{Z}_2$ invariants for a time-reversal-invariant system in 3D.  Suppose that the Brillouin zone torus is $x,y,z \in [-1,1],$ and 
%{\bf OK, at this point I do not understand where these conditions come from.}
construct an EBZ by taking the part of the torus with $z \geq 0$.  Then slices at constant $z$ for $0<z<1$ have the topology of the torus $T^2$, while at $z=0$ and $z=1$ there are additional ${\cal T}$ constraints that reduce the degrees of freedom to the 2D BZ.  The  ${\cal T}$ constraint means that the $xy$ Chern number is zero in every slice, while the $xz$ and $yz$ integers are arbitrary.
%A key feature of the 3D case is that, since the boundaries at $z=0$ and $z=1$ are equivalent to the 2D EBZ, each boundary can be in the $\mathbb{Z}_2$ ``even'' or ``odd'' class, and this is stable under deformation with the ${\cal T}$ symmetry requirements.

The EBZ boundaries at $z=0$ and $z=1$ are characterized by one $\mathbb{Z}_2$ invariant each: one boundary may be even while the other is odd because the boundary slices have the same Chern number (zero) and thus are homotopic as maps to ${\cal C}$ in the EBZ interior.
Once the two $\mathbb{Z}_2$ invariants at the boundaries are fixed, two contractions of an original 3D EBZ to the torus $T^3$ differ by two even Chern numbers, one for the $xz$ slices and one for the $yz$ slices.  These are even because a slice of the contraction has the same symmetries that force even Chern numbers in the 2D case.  There are four additive $\mathbb{Z}_2$ invariants per band pair in 3D and 16 insulating phases.

Going back to the torus $T^3$, there are 6 inequivalent planes that have the symmetries of the 2D BZ and hence have a $\mathbb{Z}_2$ invariant.   Only four of these are independent, consistent with the counting above: if the six invariants with values $\pm 1$ are $x_0, x_1, y_0, y_1, z_0, z_1$, for the planes $x=0$ and $x=\pm 1$ and so on, then there are two relations
\beq
x_0 x_1 = y_0 y_1 = z_0 z_1.
\eeq
For example, a model can be designed on the 3D sodium chloride lattice to reduce to a previously introduced~\cite{roymodel} 2D square lattice topological insulator in the $k_y=0$ or $k_z=0$ planes: it has a phase with $y_0=y_1=z_0=z_1= -1$, $x_0=x_1=1$.  In 2D, both insulating phases can be realized in models where there is a conserved quantity (e.g., $S_z$) that allows definition of ordinary Chern integers.  A new feature in 3D is that the phases with $x_0 x_1 = y_0 y_1 = z_0 z_1 = -1$ cannot be realized in this way.

Since the results here are for Hilbert spaces of arbitrary dimension, they can be applied to many-body
problems with an odd number of fermions~\cite{asss} if there are two periodic parameters
in the Hamiltonian that are connected by time-reversal in the same way
as the momentum components $(k_x,k_y)$.  Our derivation of $\mathbb{Z}_2$ invariants helps explain spin Hall edge states~\cite{km1}: in 2D, just as Chern number predicts the number of edge states in the IQHE~\cite{hatsugai}, the class of even (odd) Chern numbers in the bulk corresponds to edges with an even (odd) number of Kramers pairs of modes.  Note that transport by these edge states will receive power-law, rather than exponential, corrections in voltage or temperature.

The existence of a 2D $\mathbb{Z}_2$ invariant was also obtained by Haldane~\cite{haldaneunpub}.  The bulk-edge connection has been derived when ordinary Chern integers are defined~\cite{shenghaldane,qi,roy} (see also Ref.~\onlinecite{kaneunpub}).  Recently preprints appeared on the 2D $\mathbb{Z}_2$ invariant as an obstruction~\cite{roy} or
%, although the counting of invariants in the latter differs slightly from our results.
as a noninvariant Chern integral plus a formal integral on the EBZ boundary that is defined up to addition of an even integer~\cite{kaneunpub}.   Defining this integral is equivalent to our prescription of choosing any contraction to define a Chern integer.  To our knowledge, the full counting of $\mathbb{Z}_2$ invariants in 2D or 3D has not been obtained before.

Although whether graphene has sufficiently strong spin-orbit coupling
to realize the topological insulator has been
debated~\cite{zhangspin,macdonaldspin}, it is hoped that the
understanding of $\mathbb{Z}_2$ invariants developed here will
stimulate searches in a wider class of materials, especially in 3D.  The
understanding of bulk $\mathbb{Z}_2$ invariants in this paper,
combined with the stability of total $\mathbb{Z}_2$ to interactions
and scattering at the edge~\cite{xumooreedge,wu}, shows that the
topological insulator is a robust phase of matter with a deep
connection to the quantum Hall effect.

The authors acknowledge valuable correspondence with C.~L.~Kane and
support from NSF DMR-0238760 (J.~E.~M.), NSF DMR04-57440 (L.~B.) and
the Packard Foundation (L.~B.).

\end{document}